\documentclass[aps,prb,twocolumn,longbibliography,superscriptaddress,showpacs, 10pt]{revtex4-1}
\usepackage{graphicx}
\usepackage{dcolumn}
\usepackage{bm}
 \usepackage{tensor}
 \usepackage{physics}
\usepackage{times}
\usepackage{siunitx}

\begin{document}


\title{\vspace{-0.5 cm} \Large Realization of a scalable Laguerre-Gaussian mode sorter based on a robust radial mode sorter}


\author{Dongzhi Fu}
\affiliation{Shaanxi Key Laboratory of Quantum Information and Quantum Optoelectronic Devices, School of Science, Xi'an Jiaotong University, Xi'an 710049, China}
\affiliation{The Institute of Optics, University of Rochester, Rochester, New York 14627, USA}
\author{Yiyu Zhou}
\email{yzhou62@ur.rochester.edu }
\affiliation{The Institute of Optics, University of Rochester, Rochester, New York 14627, USA}
\author{Rui Qi}
\affiliation{The Institute of Optics, University of Rochester, Rochester, New York 14627, USA}
\author{Stone Oliver}
\affiliation{Department of Physics, Miami University, Oxford, Ohio 45056, USA}
\author{Yunlong Wang}
\affiliation{Shaanxi Key Laboratory of Quantum Information and Quantum Optoelectronic Devices, School of Science, Xi'an Jiaotong University, Xi'an 710049, China}
\author{Seyed Mohammad Hashemi Rafsanjani}
\affiliation{The Institute of Optics, University of Rochester, Rochester, New York 14627, USA}
\author{Jiapeng Zhao}
\affiliation{The Institute of Optics, University of Rochester, Rochester, New York 14627, USA}
\author{Mohammad Mirhosseini}
\affiliation{The Institute of Optics, University of Rochester, Rochester, New York 14627, USA}
\author{Zhimin Shi}
\affiliation{Department of Physics, University of South Florida, Tampa, FL 33620 USA}
\author{Pei Zhang }
\email{zhangpei@mail.ustc.edu.cn }
\affiliation{Shaanxi Key Laboratory of Quantum Information and Quantum Optoelectronic Devices, School of Science, Xi'an Jiaotong University, Xi'an 710049, China}
\author{Robert W. Boyd}
\affiliation{The Institute of Optics, University of Rochester, Rochester, New York 14627, USA}
\affiliation{Department of Physics, University of Ottawa, Ottawa, Ontario K1N 6N5, Canada}

\date{\today}

\begin{abstract}
\begin{center}
Correspondence and requests for materials should be addressed to Yiyu Zhou (yzhou62@ur.rochester.edu) and Pei Zhang (zhangpei@mail.ustc.edu.cn).
\end{center}
\vspace{0.25\baselineskip}

{\bfseries Contact details of all authors}. Dongzhi Fu, Email: fdz1010@gmail.com; Yiyu Zhou, yzhou62@ur.rochester.edu; Rui Qi, ruiqi.techdoc@gmail.com; Stone Oliver, oliversb@miamioh.edu; Yunlong Wang, 352872731@qq.com; Seyed Mohammad Hashemi Rafsanjani, mo.hashemir@gmail.com; Jiapeng Zhao, zjpapply@gmail.com; Mohammad Mirhosseini, moh.mir@gmail.com; Zhimin Shi, zshi.opt@gmail.com; Pei Zhang, zhang.pei@stu.xjtu.edu.cn; Robert W. Boyd, boydrw@mac.com.

\vspace{1\baselineskip}

{\bfseries \large Abstract}
\setlength{\parskip}{0.2\baselineskip}

 The transverse structure of light is recognized as a resource that can be used to encode information onto photons and has been shown to be useful to enhance communication capacity as well as resolve point sources in superresolution imaging. The Laguerre-Gaussian (LG) modes form a complete and orthonormal basis set and are described by a radial index $p$ and an orbital angular momentum (OAM) index $\ell$. Earlier works have shown how to build a sorter for the radial index $p$ or/and the OAM index $\ell$ of LG modes, but a scalable and dedicated LG mode sorter which simultaneous determinate $p$ and $\ell$ is immature. Here we propose and experimentally demonstrate a scheme to accomplish complete LG mode sorting, which consists of a novel, robust radial mode sorter that can be used to couple radial modes to polarizations, an $\ell$-dependent phase shifter and an OAM mode sorter. Our scheme is in principle  efficient, scalable, and crosstalk-free, and therefore has potential for applications in optical communications, quantum information technology, superresolution imaging, and fiber optics.
\vspace{-0.5 cm}
\end{abstract}

\maketitle
{\bfseries \large Introduction}
\setlength{\parskip}{0.2\baselineskip}

The transverse structure of light can be described by an infinite-dimensional Hilbert space, making it attractive for applications such as quantum information technology \cite{nagali2009quantum, krenn2014generation} and optical communications \cite{wang2011high, lei2015massive, bozinovic2013terabit, richardson2013space, wang2012terabit}. A rotationally-symmetric discrete basis set to describe the transverse field is the Laguerre-Gaussian (LG) mode set, which is characterized by two mode indices:~the radial index $p \in \{0,1,2,...\}$ and the azimuthal index $\ell \in \{0,\pm 1,\pm 2,...\}$. The azimuthal index $\ell$ is associated with a vortex phase structure  exp($i\ell\theta$) and indicates an orbital angular momentum (OAM) of $\ell\hbar$ per photon, where $\theta$ is the azimuthal angle \cite{allen1992orbital}. While the OAM modes have been broadly applied to enhancing transmission rates of both classical and quantum communications \cite{wang2011high, lei2015massive, bozinovic2013terabit, richardson2013space, wang2012terabit,molina2001management, mirhosseini2015high, sit2017high, wang2018towards,groblacher2006experimental}, it is desirable to multiplex the radial degree of freedom of LG modes since the OAM modes alone cannot reach the capacity limit of a communication channel \cite{zhao2015capacity}. Furthermore, multiplexing of both $p$ and $\ell$ has been investigated in high-dimensional entangled quantum system \cite{krenn2014generation} and shows the potential usefulness of LG modes in quantum information technology \cite{nagali2009quantum,wang201818,langford2004measuring,zhang2007demonstration, zhang2012implementing,mair2001entanglement,karimi2014radial, karimi2014exploring}. More recently, it has been demonstrated  \cite{tsang2016quantum} that spatial mode decomposition can be used to resolve closely located point sources. Instead of simply detecting the position information of photons via a camera, measuring the spatial mode components can provide higher Fisher information and beat ``Rayleigh's curse", with important implications for fluorescence microscopy and astronomy. For these reasons, it is highly desirable to develop the capability of measuring the spatial mode distribution of photons in the LG basis.

\begin{figure*}[t]
\includegraphics[width=0.6\textwidth]{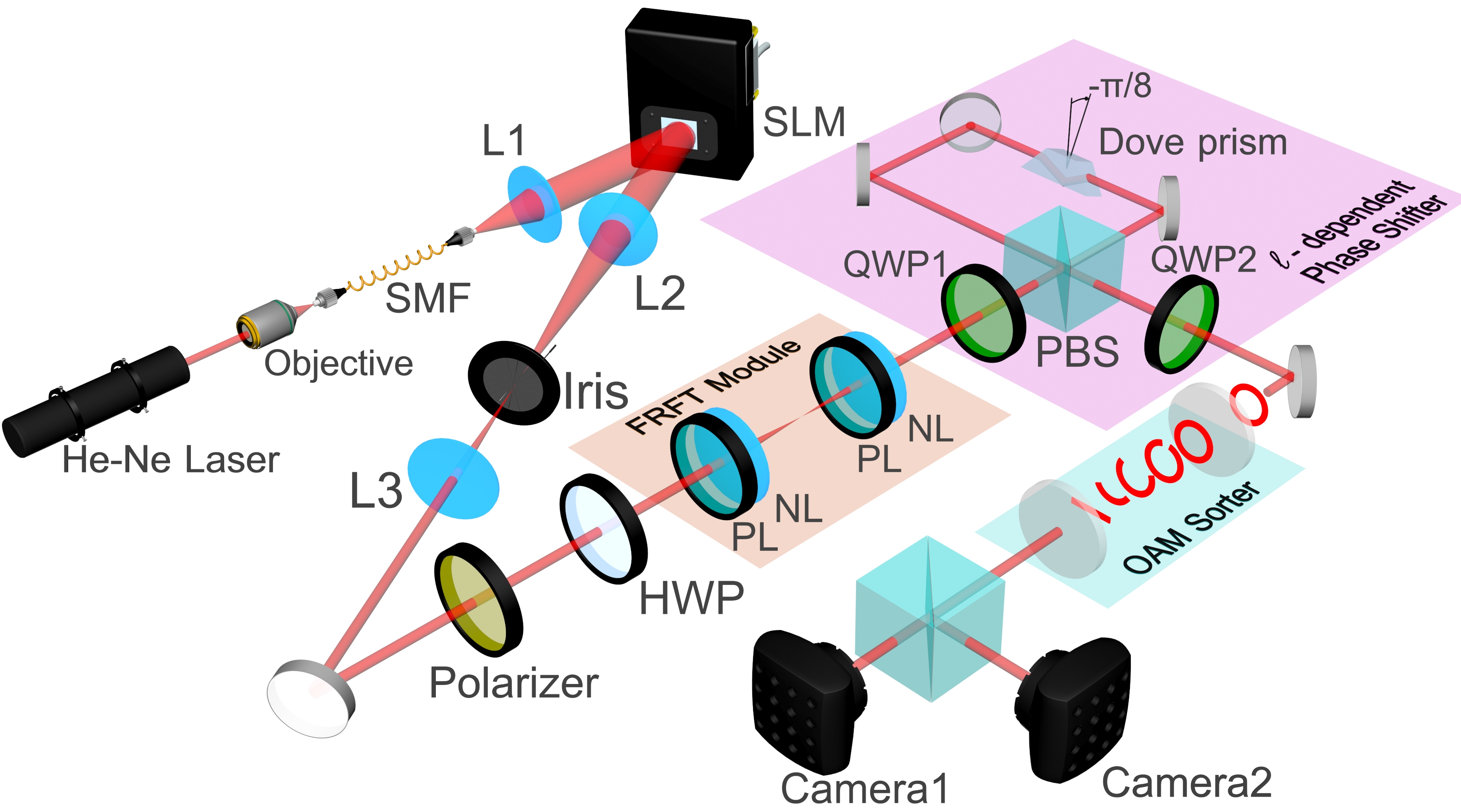}
\caption{
Experimental setup of the LG mode sorter. The LG mode is generated by a spatial light modulator (SLM). The polarizer and the half-wave plate (HWP) set the photons to be horizontally polarized. The FRFT module is used to realize a radial mode sorter. The $\ell$-dependent phase shifter is realized by a Sagnac interferometer and a Dove prism. The OAM sorter performs a coordinate transform to efficiently separate the OAM modes. SMF: single-mode
fiber; L: lens; PL:  polarization-dependent lens; NL: normal lens; HWP: half-wave plate; QWP: quarter-wave plate. }
\label{fig:modesorter}
\vspace{-0.5 cm}
\end{figure*}

For the azimuthal index $\ell$, various types of sorter have been demonstrated such as the Dove-prism-based interferometer \cite{leach2002measuring, leach2004interferometric} and the log-polar coordinate transformer \cite{mirhosseini2013efficient, berkhout2011measuring, lavery2011measurement, berkhout2010efficient,larocque2017generalized,o2012near,lavery2012refractive}. For the radial index $p$, it was recently proposed that the fractional Fourier transform (FRFT) can be used to realize a radial mode sorter to efficiently measure $p$ \cite{zhou2017sorting, gu2018gouy, zhou2018quantum}. Thus, by combining these two processes, one can obtain the full information content of the LG beam. However, for two reasons it is challenging to stably and simultaneously sort $\ell$ and $p$. First, FRFT-based radial mode sorters typically require a cascade of an unwieldy $N - 1$ Mach-Zenhder interferometers to sort N modes \cite{zhou2017sorting}, which makes it difficult to access and utilize high-order modes. Second, the FRFT-based radial mode sorter routes photons to different output ports according to the value $p+\frac{1}{2}|\ell|$. The reason for this behavior is that the Gouy phase is proportional to this factor \cite{zhou2017sorting}. Therefore, LG modes with an odd $\ell$ will be split into two output ports as a consequence of the fractional number $1/2$ before $|\ell|$. For instance, the radial mode sorter developed in ref. $32$ 
works well for $\ell=0$ and is able to separate LG mode of $p=0$, $p=1$, and $p=2$ simultaneously. However, it can be checked that LG$_{p=0}^{\ell=1}$ cannot be properly sorted and will be split to more than one output ports. Furthermore, the spatial mode demultiplexing based on multi-plane light conversion \cite{bade2018fabrication, fontaine2018optical} has been proposed several years ago and it presents a universal solution that is not dedicated to LG modes. However, small imperfections in alignment leads to significant mode cross-talks, which limits the practicality of such implementations in quantum key distribution due to the error rate threshold.  Therefore, it remains a problem to build a scalable and dedicated LG mode sorter that is able to determine $p$ and $\ell$ simultaneously and unambiguously.

\setlength{\parskip}{0.5\baselineskip}
{\bfseries \large Results}
\setlength{\parskip}{0.2\baselineskip}

\setlength{\parskip}{0.5\baselineskip}
{\bfseries \normalsize Theoretical framework}
\setlength{\parskip}{0.2\baselineskip}

Here we propose and experimentally realize a scheme to sort the radial and OAM indices of LG modes at the same time. We first implement a robust radial mode sorter by using a pair of spatially inhomogeneous half-wave plates (HWP). We refer to such a HWP as a polarization-dependent lens (PL) in that it works as a lens of opposite focal lengths $(f, -f)$ for left- and right-handed circular polarizations respectively. With this novel device, we can construct a common-path interferometer in which two circular polarizations act like two separate arms of a Mach-Zenhder interferometer, leading to a compact, robust, and cost-effective radial mode sorter. We then cascade an $\ell$-dependent phase shifter as the building block to connect the radial mode sorter and the subsequent OAM mode sorter. This $\ell$-dependent phase shifter is realized by a Sagnac interferometer and a Dove prism and can be used to counteract the $\ell$-dependent Gouy phase of FRFT as mentioned earlier. Finally we cascade an OAM mode sorter to complete the sorting of LG modes. The explicit transformation that each part of the LG mode sorter provides is given below,
\begin{equation} \label{Eq:Fulltransformations}
\begin{aligned}
&\text{LG}_p^{\ell} \otimes (\ket{L}+\ket{R}) \\
  \xrightarrow[]{\text{       FRFT       }}& \text{LG}_p^{\ell} \otimes (\ket{L}+  e^{i \pi (p+\frac{|\ell|}{2})} \ket{R}) \\
\xrightarrow[]{\text{    shifter       }} & \text{LG}_p^{\ell} \otimes (\ket{L}+  e^{i \pi (p+\frac{|\ell|}{2} +\frac{\ell}{2})} \ket{R}) \\
=&\left\{ {\begin{array}{*{20}{l}}  \text{LG}_p^{\ell} \otimes    (   \ket{L} + e^{i\pi (p+\ell)}\ket{R}  )   ,  \quad   \ell	> 0\\
\text{LG}_p^{\ell} \otimes (   \ket{L} +e^{i\pi  p }\ket{R}  )   ,  \quad \ell 	\leq 0 \\
\end{array}.} \right.
\end{aligned}
\end{equation}
The detailed analysis of each component and the derivation of Eq.~(\ref{Eq:Fulltransformations}) are presented in Materials and methods

\setlength{\parskip}{0.5\baselineskip}
{\bfseries \normalsize Experimental setup}
\setlength{\parskip}{0.2\baselineskip}

The experimental setup of the LG mode sorter is presented in Fig.~\ref{fig:modesorter}. A 633 nm HeNe laser is coupled to a single-mode fiber (SMF) and is then collimated to illuminate a spatial light modulator (SLM). A binary computer-generated hologram is imprinted on the SLM to generate the desired LG mode at the first diffraction order \cite{mirhosseini2013rapid}. The generated mode becomes horizontally polarized by using a polarizer and a HWP and is then injected into a FRFT module. The FRFT module consists of two sets of lenses, and each set has a normal lens (NL) and a PL (Edmund Optics \#34-466) placed together. The experimental parameters used to realize FRFT are calculated according to Eq.~(3) in ref. $32$. The $\ell$-dependent phase shifter consists of a Sagnac interferometer and a Dove prism and is cascaded to the FRFT module to remove the fractional number $1/2$ that appears in Eq.~(\ref{Eq:Fulltransformations}). Then we cascade a polarization-independent OAM sorter based on log-polar coordinate transformation \cite{mirhosseini2013efficient, berkhout2011measuring, lavery2011measurement, berkhout2010efficient} to the $\ell$-dependent phase shifter. Finally, we put two cameras to record the output images at the two output ports.

\setlength{\parskip}{0.5\baselineskip}
{\bfseries \normalsize Experimental demonstration}
\setlength{\parskip}{0.2\baselineskip}

\begin{figure}[t!]
\includegraphics[width=\columnwidth]{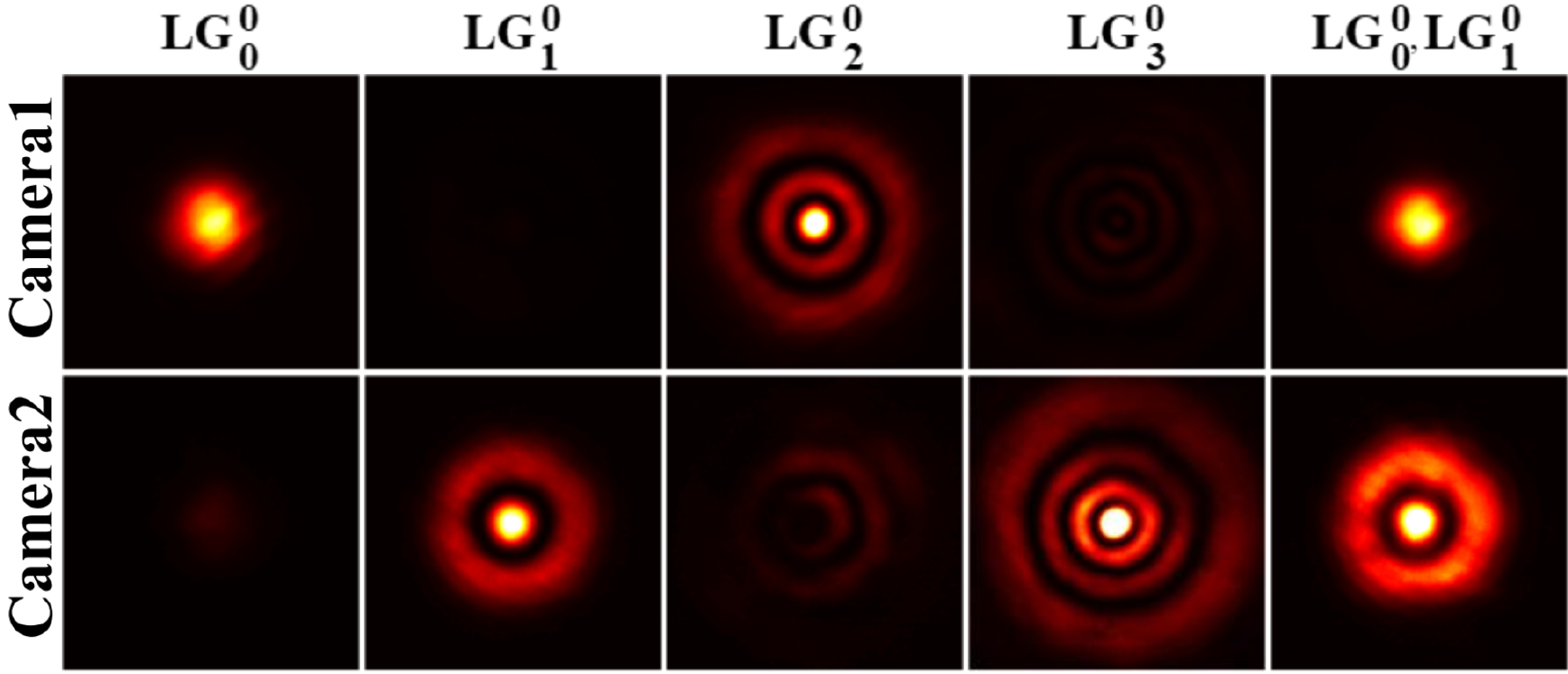}
\caption{
Experimental results for the radial mode sorter when $\ell$ of the input LG modes is $0$. The first four columns show that a LG mode of odd (even) $p$ is sorted to Camera2 (Camera1). The last column shows the result when the input is a coherent superposition state of $p=0$ and $p=1$.}
\label{fig:Presults}
\vspace{-0.5 cm}
\end{figure}

The experimental results of our radial mode sorter for the modes of $\ell=0$ is shown in Fig.~\ref{fig:Presults}, which is obtained by using a PBS and two cameras immediately after the FRFT module. According to Eq.~(\ref{Eq:Fulltransformations}), when $\ell$ is zero, an odd (even) radial index $p$ corresponds to a vertical (horizontal) polarization and can be sorted to Camera2 (Camera1) by a PBS as shown in Fig.~\ref{fig:Presults}. Moreover, the radial mode sorter also works for the superposition states, as shown in the last column of Fig.~\ref{fig:Presults}. An equal superposition state of LG$_{p=0}^{\ell=0}$ and LG$_{p=1}^{\ell=0}$ is injected, and LG$_{p=0}^{\ell=0}$ and LG$_{p=1}^{\ell=0}$ modes are sorted to Camera1 and Camera2 respectively. We can also experimentally characterize the crosstalk for this radial mode sorter, which is defined to be the power in the wrong port divided by the total output power. In our experiment, the measured crosstalk for LG$_{p=0}^{\ell=0}$ and LG$_{p=1}^{\ell=0}$ are $8.63\%$ and $7.50\%$, respectively. In addition we emphasize that this crosstalk results from the misalignment and experimental imperfections. We believe that using high-quality antireflection coated optics, matching focal length between NL and PL, accurate distant in FRFT module and more careful alignment can mitigate crosstalk significantly and bring the radial mode sorter to its theoretical limit of $100\%$ efficiency and no crosstalk.

\begin{figure}[t!]
\includegraphics[width=\columnwidth]{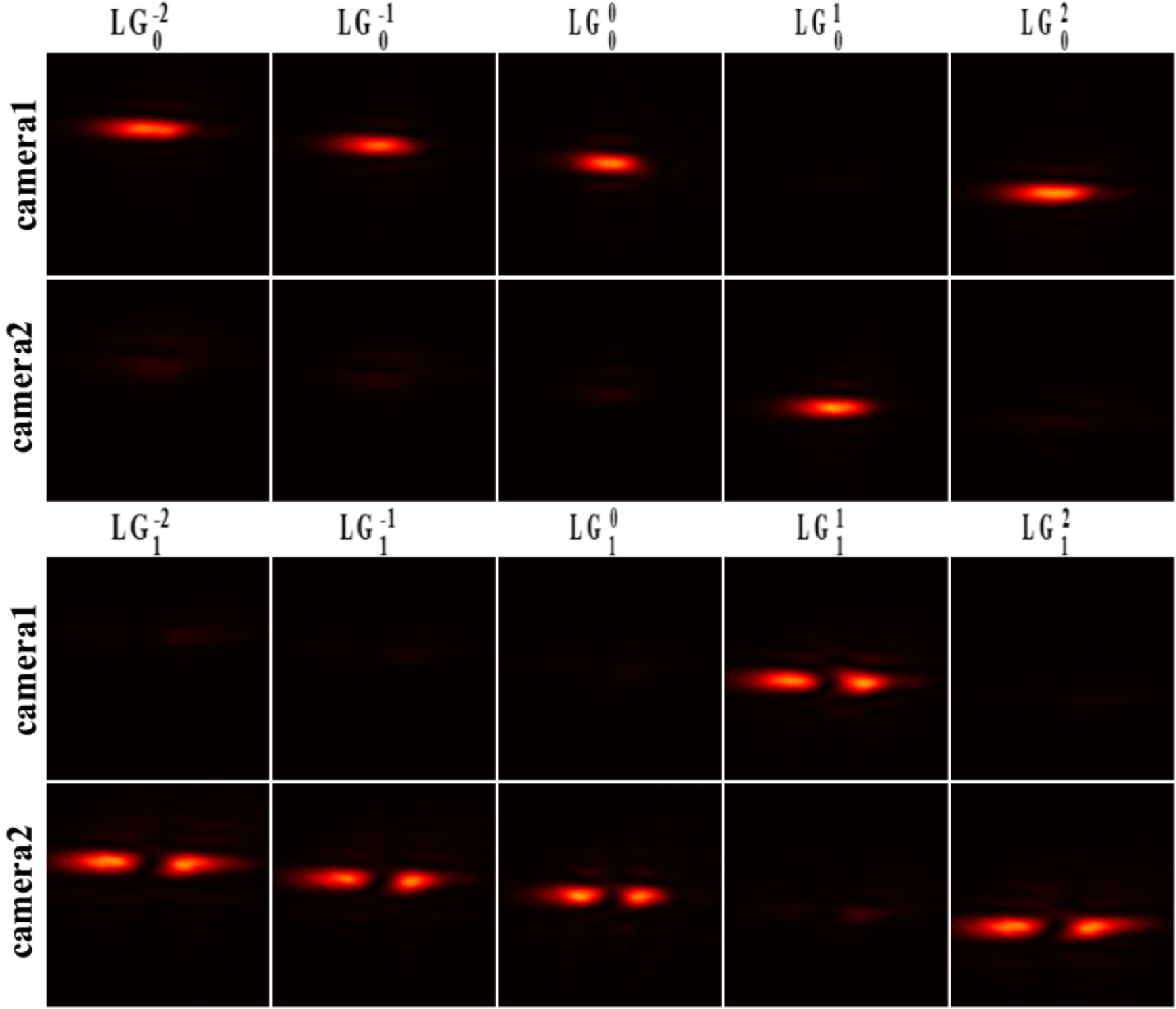}
\caption{
Experimental results for the LG mode sorter. The input state is shown on the top of the images. When $\ell$ is non-positive, it can be seen that $p=0$ and $p=1$ modes are sorted to Camera1 and Camera2 respectively, and the $\ell$ value determines the vertical position of sorted modes. When $\ell$ is positive, the LG modes with an odd (even) value of $p+\ell$ are directed to Camera2 (Camera1), while the vertical position of sorted mode is still determined by $\ell$.}
\label{fig:Lsorter}
\vspace{-0.5 cm}
\end{figure}

\begin{figure}[t]
\includegraphics[width=0.85\columnwidth]{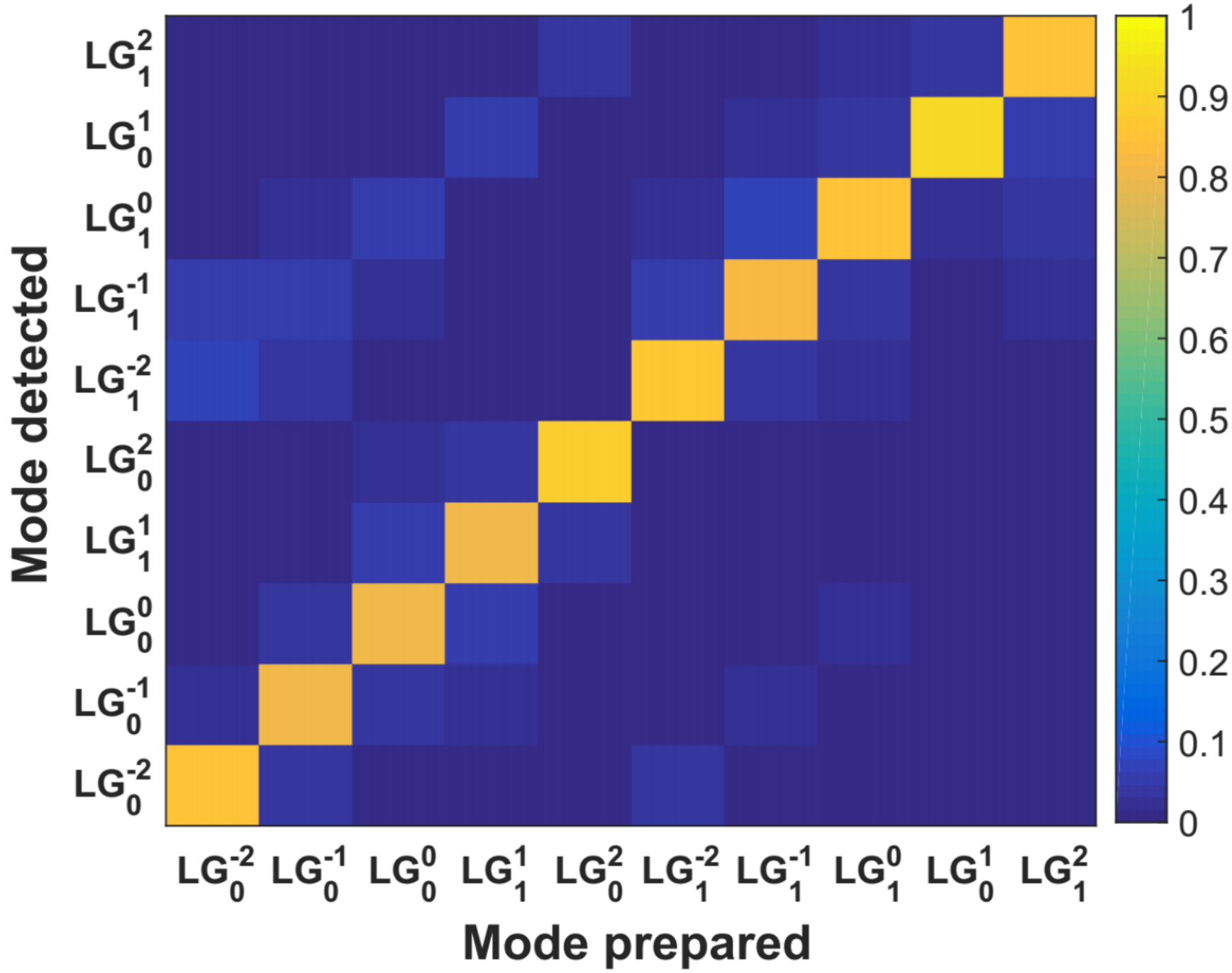}
\caption{
The measured crosstalk matrix of the LG mode sorter.}
\label{fig:crosstalk}
\vspace{-0.5 cm}
\end{figure}

With the help of the subsequent $\ell$-dependent phase shifter and the OAM sorter, the LG modes are mapped onto different positions in different output ports, as shown in Fig.~\ref{fig:Lsorter}. For a non-positive $\ell$, all modes with an even $p$ appear in Camera1, but all modes with an odd $p$ appear in Camera2. In other words, $p$ determines which camera the mode will be sorted into, and $\ell$ determines the vertical position on the camera. For a positive $\ell$, all modes with an even (odd) value of $p+\ell$ appear in Camera1 (Camera2). That is, for a given value of $p$, the neighboring $\ell$ modes appear not only in different vertical positions but also in different cameras. Despite of this complication, it can be seen that different LG modes are mapped to a unique position and thus the setup is essentially an effective sorter. Additionally, after the log-polar transformation, the intensity distribution along the horizontal axis in Fig.~\ref{fig:Lsorter} corresponds to the radial profile of the incident LG modes, therefore more than one spot can be seen for $p\geq1$. The experimental crosstalk matrix of the LG mode sorter is shown in Fig.~\ref{fig:crosstalk}, and the average crosstalk is 15.3\%. We note that because of the insufficient mode resolution of our OAM sorter, we reduce the detection area on the camera for each sorted mode to decrease the crosstalk, and this leads to a loss of 39.1\% on average. However, we emphasize that this sorter still substantially outperforms methods base on projective measurement, because the efficiency of projective measurement is bounded to $1/d$ for a $d$-dimensional state space, which results in a loss of at least 90\% for these 10 LG modes presented here. In addition, the mode resolution of the OAM sorter can be readily enhanced by adding a beam-copying grating \cite{mirhosseini2013efficient}. In principle, our proposed method can have $100\%$ efficiency with intrinsically zero crosstalk.

\setlength{\parskip}{0.5\baselineskip}
{\bfseries \large Discussion}
\setlength{\parskip}{0.2\baselineskip}

\begin{figure*}[t]
\includegraphics[width=0.75\textwidth]{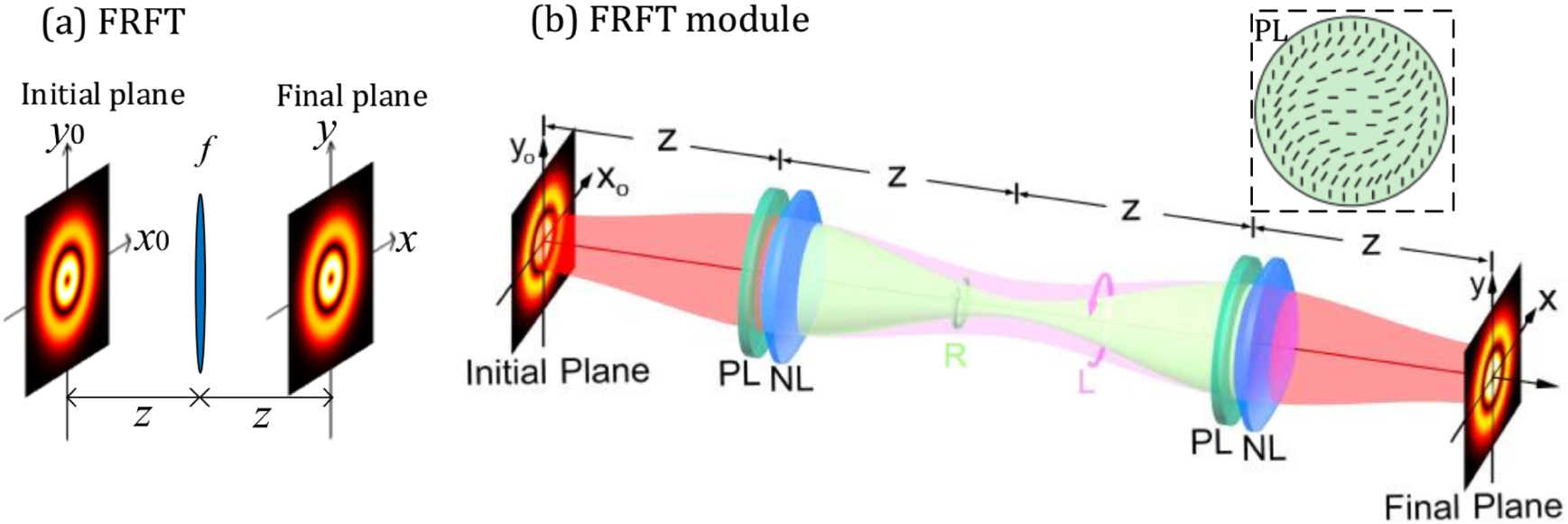}
\caption{
(a) Implementation of the FRFT with a lens. The LG mode keeps invariant in the final plane because it is the eigenmode of FRFT. Here, a mode with $p=1, \ell=1$ is presented as an example. (b) Schematic diagram of a FRFT module. The FRFT module consists of two sets of lenses (each set of lenses is combined a PL with a normal lens (NL)) and can perform a FRFT of order $\pi/2$ and $\pi$ to right-hand ($R$) and left-handed ($L$) circular polarization respectively. The inset shows an example of a PL. The black lines denote the direction of the fast axis.}
\label{fig:PLFRFT}
\vspace{-0.5 cm}
\end{figure*}

The PL we used is analogous to the q-plate \cite{marrucci2006optical} and can be used to couple radial modes to polarizations. By taking advantage of a PL, we demonstrate a robust, common-path FRFT module that can be used to identify the radial index $p$. Note that our radial mode sorter can sort $p$ modes at the single-photon level with a theoretical efficiency of $100\%$ and preserve the original radial modes. We also insert an $\ell$-dependent phase shifter to connect the FRFT module and the subsequent OAM sorter so that all LG modes can be well separated. However, we note that the sorting capability of our radial mode sorter is constrained by the FRFT module. To further increase the dimension of the radial mode sorter, we can generalize the scheme and cascade more FRFT modules with appropriate orders \cite{zhou2017sorting}. It should be noticed that the corresponding $\ell$-dependent phase shifter also needs to be adjusted accordingly. The FRFT order and the orientation angle of the Dove prism in the next stage should be half of that of the previous stage. Cascading $N-1$ FRFT modules allows the sorting of $N$ radial indices. We note that the OAM mode sorter should be moved to the end to sort the azimuthal index $\ell$. It is straightforward to further increase the dimension of our LG mode sorter by cascading more FRFT modules and $\ell$-dependent phase shifters with appropriate settings. Our scheme is able to determine $p$ and $\ell$ simultaneously, and in principle can work at the single photon level, and therefore provides a valuable access to an infinite-dimensional Hilbert space of LG modes. We expect our scheme to be useful for quantum state tomography \cite{leonhardt1996discrete}, quantum communication \cite{mirhosseini2015high}, and quantum computation \cite{nielsen2002quantum}.

\setlength{\parskip}{0.5\baselineskip}
{\bfseries \large Materials and methods}
\setlength{\parskip}{0.2\baselineskip}

The FRFT is a generalization of the Fourier transform, and the LG modes remain invariant under such a transformation except for a mode-dependent fractional Gouy phase, which can be expressed as  \cite{zhou2017sorting, almeida1994fractional}
\begin{equation}\label{Eq:eigenmodes}
{{\cal F}^a}[\text{LG}_p^\ell({r_0},{\theta _0})] =\exp[- i2a\cdot (p +|\ell|/2 )] \cdot \text{LG}_p^\ell(r,\theta ),
\end{equation}
where $a$ is the order of the FRFT, and $\text{LG}_p^\ell$ represents a LG mode of radial index $p$ and azimuthal index $\ell$. The implementation of FRFT is similar to that of a Fourier transform, as shown in Fig.~\ref{fig:PLFRFT} (a) \cite{lohmann1993image}. The propagation distance $z$ and the focal length of the lens $f$ are related to the FRFT order $a$, the wavelength $\lambda$, and the beam waist radius $w_0$ through the following equations \cite{zhou2017sorting}:
\begin{align}\label{Eq:parameters}
z = \frac{\pi w_0^2}{\lambda}\tan\frac{a}{2}, ~~~~~~~~~
f = \frac{\pi w_0^2}{\lambda \sin a}.
\end{align}
Given the $p$-dependent Gouy phase $- 2a\cdot (p +|\ell|/2 )$, we can perform FRFT of different orders in two arms of a Mach-Zenhder interferometer to realize a radial mode sorter \cite{zhou2017sorting}. To improve the stability, we employ two orthogonal polarizations as two arms in the interferometer to build an inherently stable, common-path radial mode sorter \cite{zhou2018quantum}. Such a configuration requires a polarization-dependent FRFT, which can be achieved by a PL as shown in Fig.~\ref{fig:PLFRFT} (b). The PL is a spatially inhomogeneous HWP whose fast-axis angle has a spatial dependence and can be used to couple radial modes to polarizations. A widely used example is the $q$-plate whose fast-axis angle is a function of azimuthal angle and can be used to realize spin-to-orbital angular momentum conversion \cite{marrucci2006optical}. Here we consider a radially varying inhomogeneity in which the fast-axis angle $\alpha$ is a function of radius and can be expressed as
\begin{equation}\label{Eq:PLAngle}
\alpha (r)= \frac{{\pi {r^2}}}{{2\lambda {f_{0}}}},
\end{equation}
where $r$ is the radial coordinate,  $\lambda$ is the wavelength, and $f_{0}$ is the effective focal length. The effect of such a PL is analyzed as follows. Assume the input optical field is represented as
\begin{equation}\label{Eq:input}
\ket{\Phi}_{\text{in}}  = E_{L} \ket{L} +  E_{R} \ket{R},
\end{equation}
where $\ket{L}$ and $\ket{R}$ are the left- and right-handed circular polarization states expressed in Dirac notation, and $E_{L}$ and $E_{R}$ are the corresponding field amplitudes. Using Jones calculus, one can readily verify that the output field of the PL becomes
\begin{equation}\label{Eq:output}
\begin{split}
\ket{\Phi}_{\text{out}}   =E_{L}     e^{i2\alpha(r)}     \ket{R}   +    E_{R}       e^{-i2\alpha(r)}        \ket{L}  .
\end{split}
\end{equation}
It can be noticed that the handedness of the circular polarization will be flipped and two conjugate phases will be impressed on the two circular polarizations. For an input $\ket{L}$ and $\ket{R}$, the PL works as a lens of focal length $-f_{0}$ and $f_{0}$, respectively. The polarization-dependent focal length makes such a device ideal for constructing a common-path FRFT module, which operates FRFTs of different orders on the two circular polarizations and can serve as an essential part of the radial mode sorter.

The realization of the FRFT module based on the PL is presented in Fig.~\ref{fig:PLFRFT} (b). The entire setup consists of two sets of lenses and can perform a FRFT of order $\pi$ for left-handed circular polarization and $\pi/2$ for right-handed circular polarization. Note that two PLs in the FRFT module are set to be back-to-back because the PL can flip polarization handedness, and we note that after two PLs the circular polarizations remain the same. It can be checked that the $\ket{L}$ simply goes through a 4-$f$ imaging system while $\ket{R}$ undergoes a Fourier transform, and a relative phase of $ \exp[i \pi (p+|\ell|/2)]$ is introduced between two circular polarizations. The polarization of the input LG mode is set to be horizontal, which forms an equal superposition of two circular polarizations and can be written as $\text{LG}_p^{\ell} \otimes (\ket{L}+\ket{R})$. Here the normalization factor is neglected for simplicity. This FRFT module will impress different phase shifts onto output beams of different circular polarizations according to their LG mode indices, and thus the following transformation can be achieved:
\begin{equation} \label{Eq:transformation}
\text{LG}_p^{\ell} \otimes (\ket{L}+\ket{R}) \rightarrow \text{LG}_p^{\ell} \otimes (\ket{L}+  e^{i \pi (p+\frac{|\ell|}{2})} \ket{R}).
\end{equation}
It can be noted that, for a fixed $\ell$, the radial index $p$ can change the output polarization, and thus it allows us to sort the radial modes by using a polarizing beamsplitter \cite{zhou2018quantum}.

However, the FRFT-based radial mode sorter does not work appropriately in some cases because of the coefficient $1/2$ before $|\ell|$ in Eq.~(\ref{Eq:transformation}) as explained earlier. For example, LG modes with an even $\ell$ will become horizontally or vertically polarized after the FRFT module and can be well separated by a PBS, but LG modes with an odd $\ell$ will become anti-diagonally or diagonally polarized and will be split to both output ports of the PBS setting in horizontal and vertical directions. In other words, the radial mode sorter can sort modes of even $\ell$ or odd $\ell$ but cannot do both simultaneously. In order to overcome this limitation, here we use an $\ell$-dependent phase shifter consisting of a Dove prism and a Sagnac interferometer as shown in  Fig.~\ref{fig:modesorter}. When the Dove prism is rotated at an angle of $\beta$, an $\ell$-dependent phase $ \exp (\pm 2i\ell\beta)$ is introduced in each arm, where the sign of the phase depends on the incident circular polarization \cite{leach2002measuring, leach2004interferometric}. Two quarter-wave plates (QWPs) before and after the Sagnac interferometer are used to realize the conversion between circular polarizations and linear polarizations. Therefore, after passing through the $\ell$-dependent phase shifter, a phase of $\exp(4i\ell\beta)$ will be introduced between $\ket{R}$ and $\ket{L}$. In our experiment we cascade the $\ell$-dependent phase shifter with the FRFT module and set $\beta = \pi/8$, and hence a relative phase of exp($i\pi \ell /2$) is introduced between two circular polarizations and the state evolution can be written as
\begin{equation} \label{Eq:Fulltransformation}
\begin{aligned}
& \text{LG}_p^{\ell} \otimes (\ket{L}+  e^{i \pi (p+\frac{|\ell|}{2})} \ket{R}) \\
\xrightarrow[]{\text{  shifter }} & \text{LG}_p^{\ell} \otimes (\ket{L}+  e^{i \pi (p+\frac{|\ell|}{2} +\frac{\ell}{2})} \ket{R}) \\
=&\left\{ {\begin{array}{*{20}{l}}  \text{LG}_p^{\ell} \otimes    (   \ket{L} + e^{i\pi (p+\ell)}\ket{R}  )   ,  \quad   \ell	> 0\\
\text{LG}_p^{\ell} \otimes (   \ket{L} +e^{i\pi  p }\ket{R}  )   ,  \quad \ell 	\leq 0 \\
\end{array}.} \right.
\end{aligned}
\end{equation}
Thus, the $\ell$-dependent phase shifter removes the fractional number 1/2 that appears in the FRFT and therefore all LG modes can be well separated by the FRFT module. Moreover, the value of $\ell$ can be obtained by cascading an OAM sorter to the output of the $\ell$-dependent phase shifter, and in this way both $p$ and $\ell$ can be obtained unambiguously. However, the common-path radial mode sorter is bounded to a two-dimensional output space, and the sorting capability is thus constrained. For example, $p=0$ and $p=1$ modes can be well separated, but $p=2$ mode will be sorted to the same position as the $p=0$ mode. We note that this problem can be removed by cascading more FRFT modules to increase the sorting capability as mentioned earlier. What's more, the LG mode sorter is effective to sort the superposition states with the help of an extra unitary transform.
%


\setlength{\parskip}{0.5\baselineskip}
{\bfseries \large Acknowledgments}
\setlength{\parskip}{0.2\baselineskip}

This work is supported by the U.S. Office of Naval Research, the Natural Science Foundation of Shaanxi Province (Grant No. 2017JM6011), and National Natural Science Foundation of China (Grants No. 91736104, 11374008, and 11534008). In addition, R. W. B. acknowledges support from Canada Research Chairs Program, the National Science and Engineering Research Council of Canada, and the Canada First Research Excellence Fund. D. F. acknowledges support from China Scholarship Council overseas scholarship. We thank M.J. Padgett and M.P. Lavery for their help in the azimuthal mode sorting.

\setlength{\parskip}{0.5\baselineskip}
{\bfseries \large Author contributions}
\setlength{\parskip}{0.2\baselineskip}

D.F. and Y.Z. initiated the project and conceived the idea. D.F., R.Q., S.O. and Y.Z. performed the experiment. D.F. and Y.Z. analyzed the results and wrote the manuscript with assistance from all authors. P.Z. and R.W.B. supervised the project.

\setlength{\parskip}{0.5\baselineskip}
{\bfseries \large Conflict of interest}
\setlength{\parskip}{0.2\baselineskip}

The authors declare that they have no conflict of interest.


\bibliography{RealizationofascalableLaguerre-Gaussianmodesorterbasedonarobustradialmodesorter}


\begin{figure*}[t]
\includegraphics[width=0.75\textwidth]{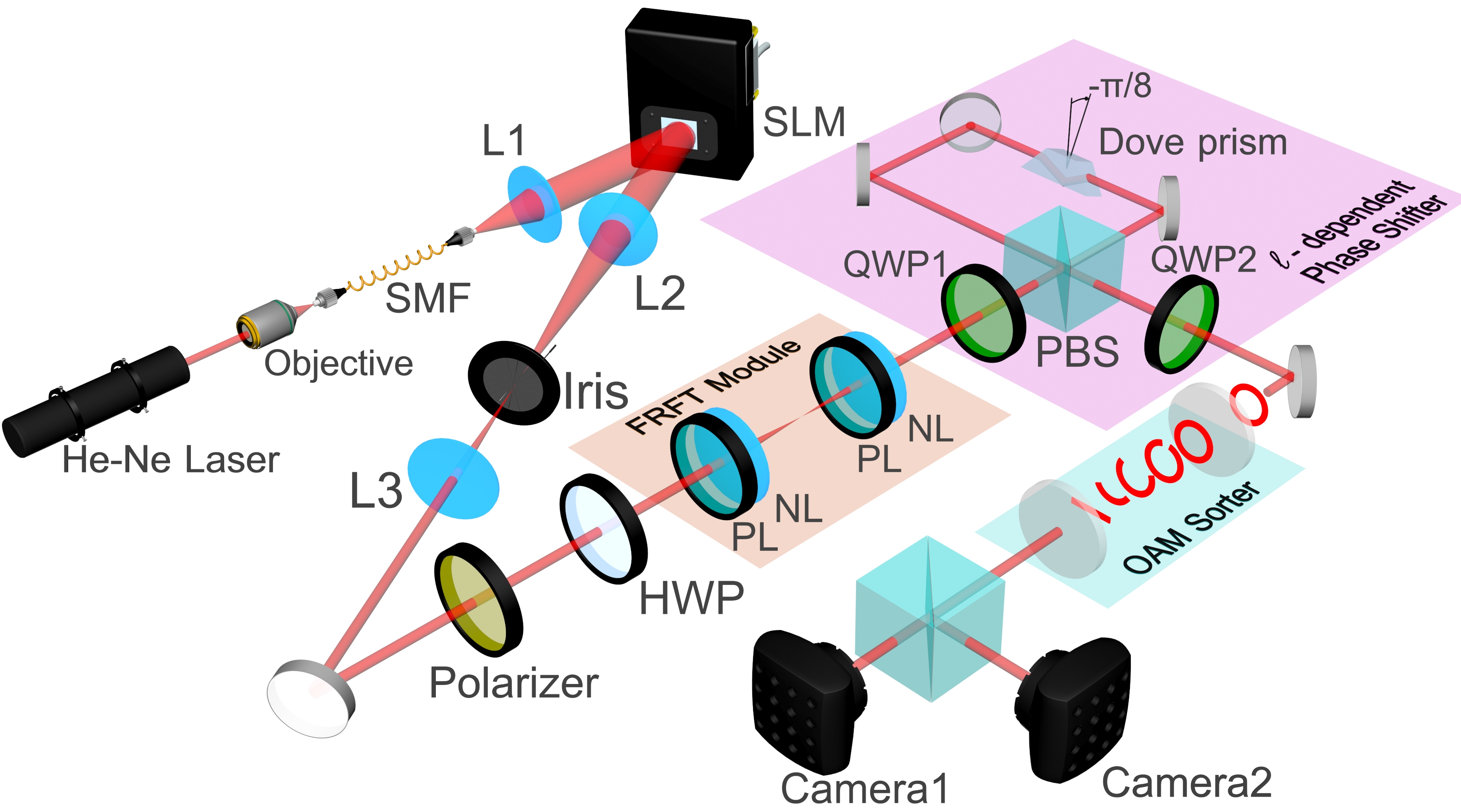}
\vspace{-0.5 cm}
\end{figure*}

\begin{figure*}[t]
\includegraphics[width=0.75\textwidth]{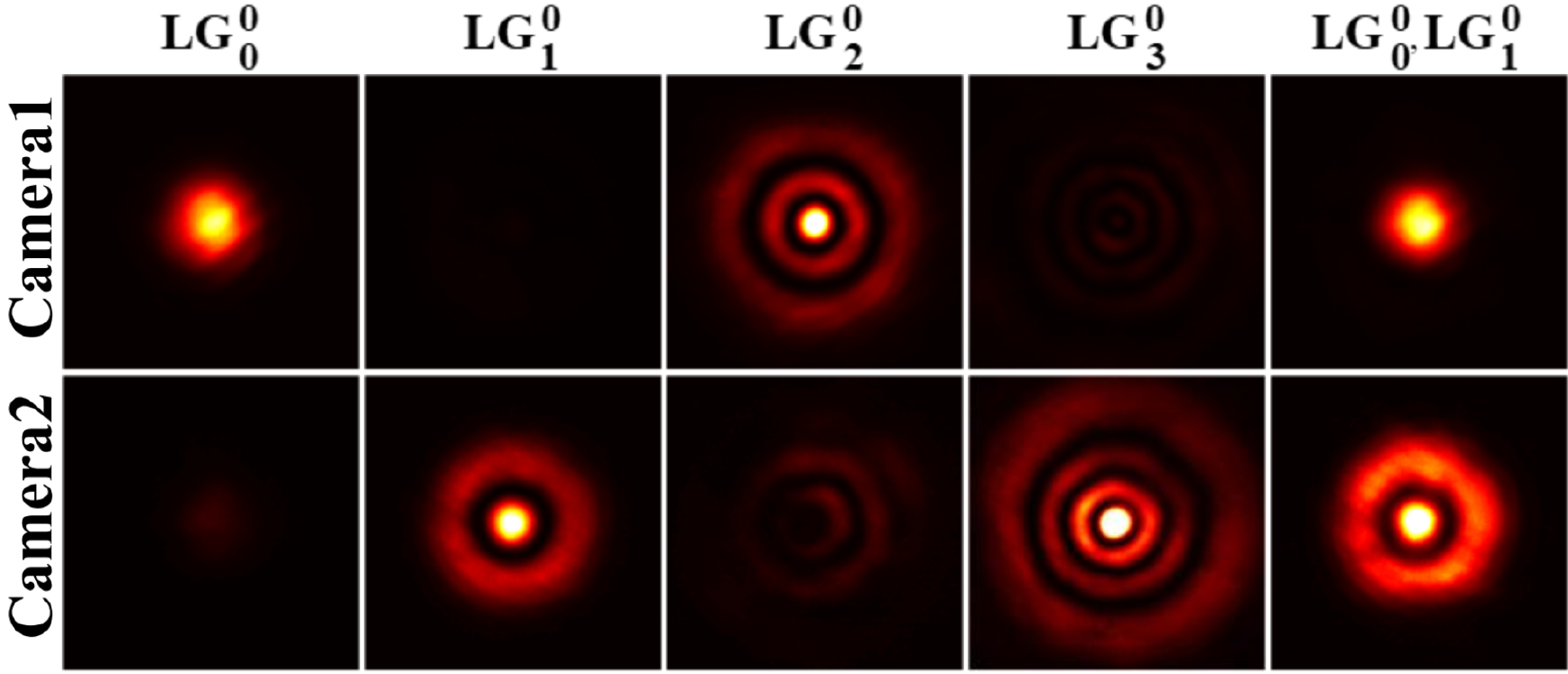}
\vspace{-0.5 cm}
\end{figure*}

\begin{figure*}[t]
\includegraphics[width=0.75\textwidth]{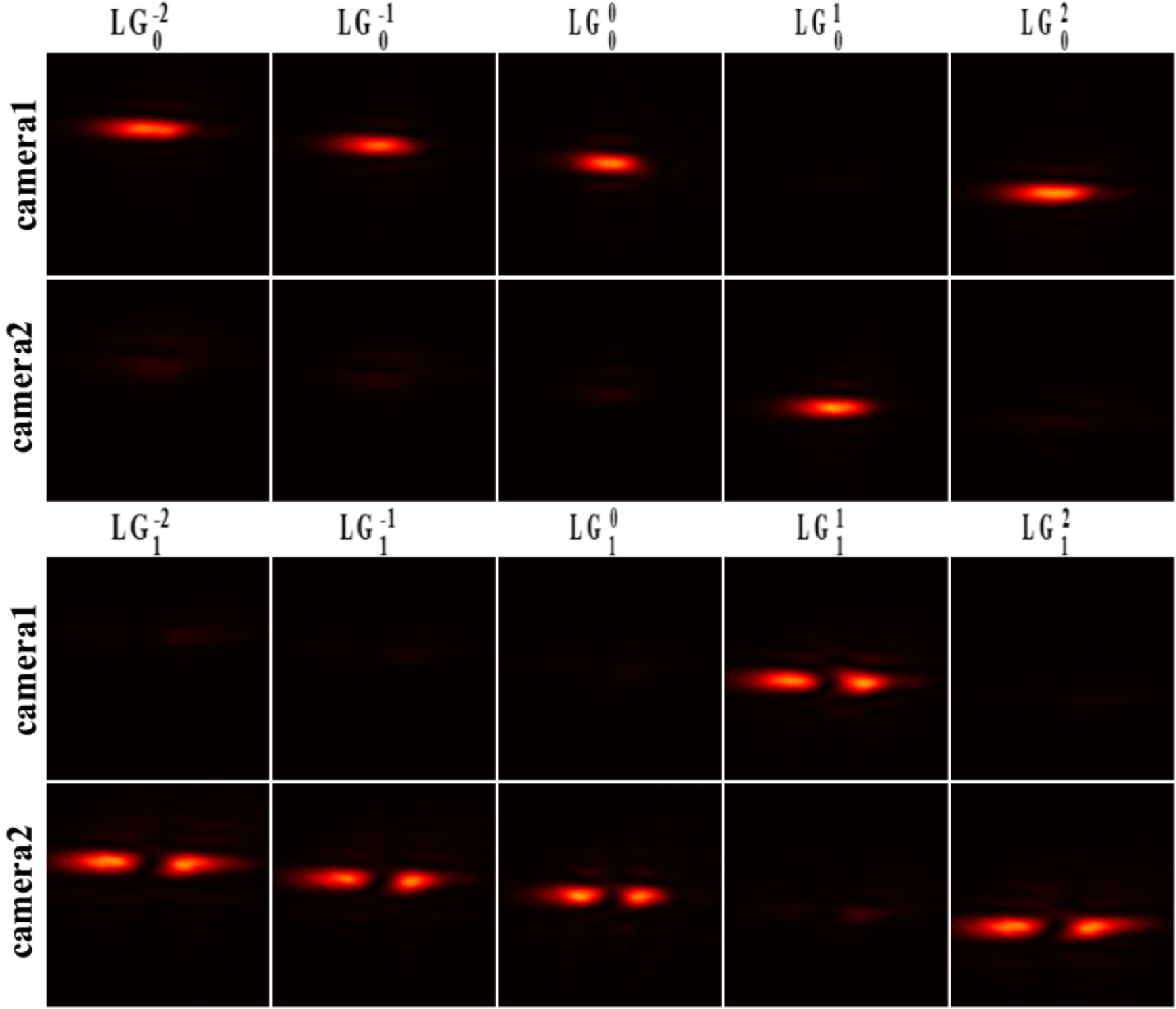}
\vspace{-0.5 cm}
\end{figure*}

\begin{figure*}[t]
\includegraphics[width=0.75\textwidth]{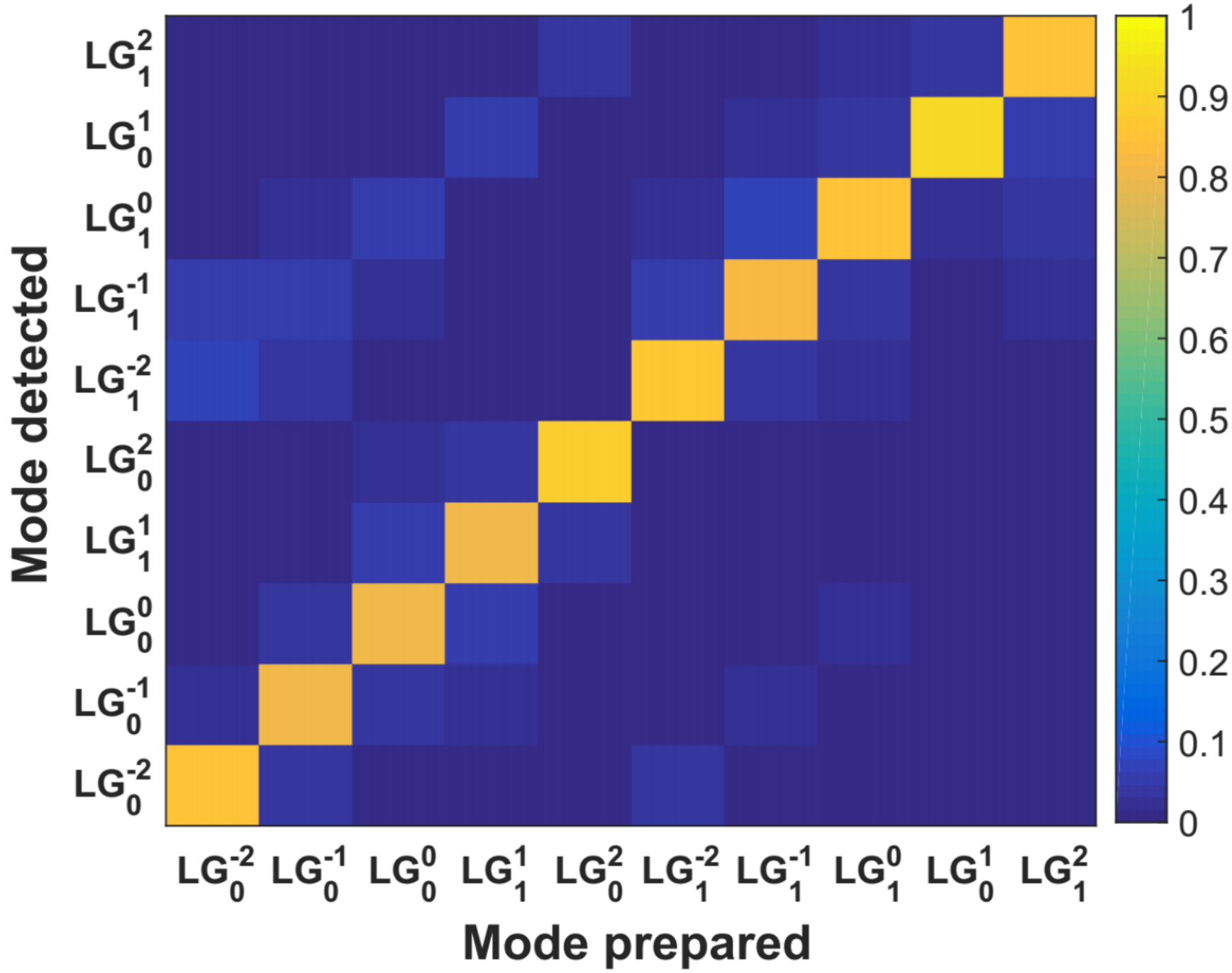}
\vspace{-0.5 cm}
\end{figure*}

\begin{figure*}[t]
\includegraphics[width=0.75\textwidth]{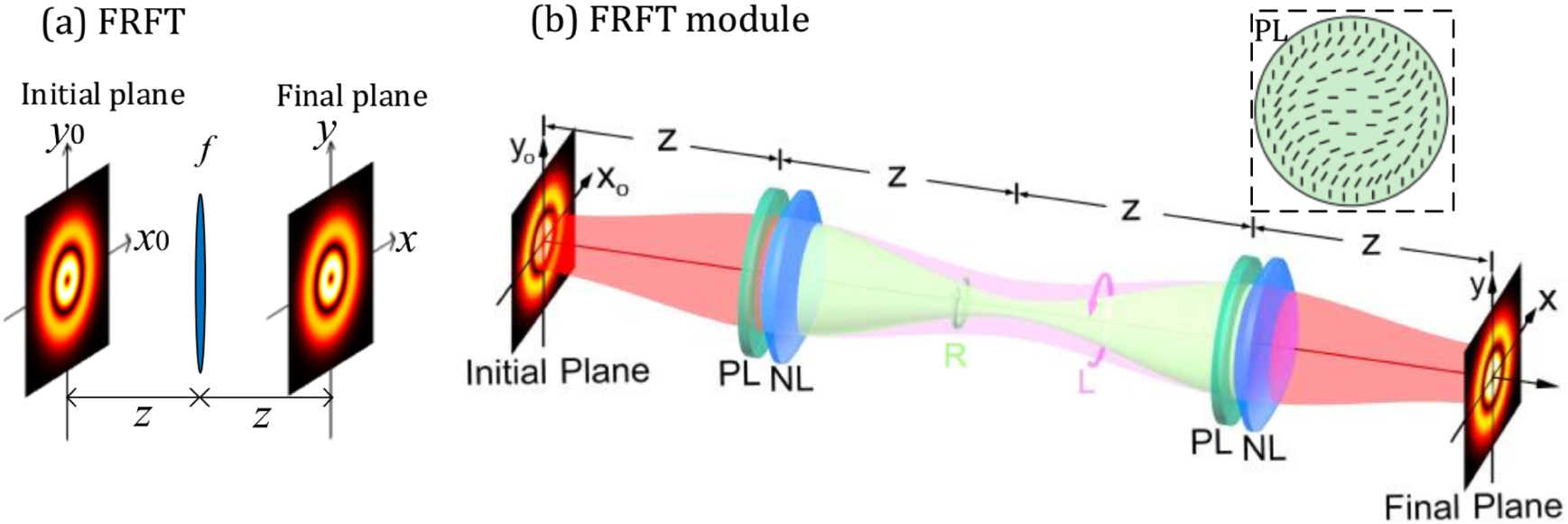}
\vspace{-0.5 cm}
\end{figure*}

\end{document}